\documentstyle[aps]{revtex} 
\begin{document} 
\draft 
\title{Four-point Green functions in the Schwinger Model} 
\author{Tomasz Rado\.zycki\thanks{Elecronic mail:
torado@fuw.edu.pl} and J\'ozef M. Namys{\l}owski\thanks{Electronic mail: 
jmn@fuw.edu.pl}} 
\address{Physics Department, Warsaw University, ul. Ho\.za 
69, 00-681 Warsaw, Poland} 
\date{\today} 
\maketitle 
\begin{abstract} 
The evaluation of the 4-point Green functions in the 1+1 Schwinger model
is presented both in momentum and coordinate space representations. The
crucial role in our calculations play two Ward identities: i) the
standard one, and ii) the chiral one. We demonstrate how the infinite set
of Dyson-Schwinger equations is simplified, and is so reduced, that a
given n-point Green function is expressed only through itself and 
lower ones. For the 4-point Green function, with two bosonic and two
fermionic external `legs', a compact solution is given both in momentum and
coordinate space representations.  For the 4-fermion Green function a
selfconsistent equation is written down in the momentum representation and
a concrete solution is given in the coordinate space. This exact solution
is further analyzed and we show that it contains a pole corresponding to
the Schwinger boson.  All detailed considerations given for various
4-point Green functions are easily generizable to higher functions.
\end{abstract} 
\section{Introduction} 
\label{sec:intro} 
 
Massless quantum electrodynamics in 1+1 space-time dimensions, known as
the Schwinger~\cite{jsch} Model (SM), proved to be a very fruitful example
of quantum field theory. Thanks to its symmetries it is a completely
solvable model, and therefore it is particularly well suited for studying
nonperturbative effects.  
 
One of the most important and well known observations is that, the
initially massless boson, called hereafter a `photon' (if one may say of
`photons' in two dimensions), aquires a mass --- the so called Schwinger
mass. As the consequence of this, the electromagnetic potential becomes a
function exponentially decreasing in space and proportional to $e^{-\mu
|x|}$, where $\mu =e / \sqrt{ \pi}$ represents the Schwinger mass of the
dressed photon. Responsible for this effect is the vacuum polarization
which totally shields the charge~\cite{cks}.  The effect of charge
screening is also known from perturbative calculations in the ordinary,
4-dimensional QED, giving rise to a weak deviation from the Coulomb law,
particularly for small distances~\cite{iz}, while in SM the change is
dramatic. The interpretation of this massive state (composite vs.
elementary) depends on the particular field variables chosen to describe
the model.

The photon mass generation mechanism appears already on the diagramatical 
level, because the exact (nonperturbative) vacuum polarization 
scalar $\Pi(k^2)$ posesses a first order pole at $k^2=0$, with the 
residuum equal to $\mu ^2$. This is commonly known as the {\it Schwinger 
mechanism}. On the other hand, from the mathematical point of view, the 
nonzero photon mass results in SM from the noninvariance of the path integral 
fermion measure with respect to the local chiral gauge transformations 
--- the $U_A(1)$ group --- which in turn is a reflection of the 
presence of anomaly in the model~\cite{fuji,rosk,sara,abh}. 

This vector meson mass generation through screening effects, is also of
interest in electroweak theory, where the additional, and still
unobserved, Higgs field has to be introduced ``by hand'', to ensure the
simultaneous renormalizability of theory and the nonzero masses of the
intermediate bosons $W^{\pm}$ and $Z^0$. 
 
Another important property of SM is the absence of the asymptotic
fermionic states~\cite{cks}. This in turn is interesting from the point of
view of hadron structure investigations~\cite{stin}, where the permanent
quark confinement and asymptotic freedom of QCD, giving rise to the
nonperturbative mass scale $\Lambda_{QCD}$, as the necessary mathematical
ingredient of the logarithmic fall off, also precludes the appearence of
the asymptotic quark states. Yet the other similarity between SM and QCD
is the existence of a fermion condensate~\cite{lowe,jaye,cad}, though this
requires considering a nontrivial instanton sector.  The above features of
SM are also preserved in a generalization of SM, by allowing fermions to
have a nonzero mass~\cite{cjs,ad1}.
 
Thanks to its full solvability, the SM, on an equal footing with other
models, as for instance the Thirring Model~\cite{thir}, may also be used
to test various assumptions in the nonperturbative calculations in quantum
field theories. As examples can serve here: i) the postulated infrared
form of the vertex function in the 4-dimensional massive
electrodynamics~\cite{sal}, applied later in the so called {\it Gauge
Technique}~\cite{gt} and other works in the context of nonperturabative
solutions of QED (for which the transverse corrections may be found in SM
explicitely~\cite{stam}), ii) the renormalisation group
methods~\cite{yild}, or iii) even the very formulation of the quantum
field theory~\cite{lowe,wotz,schm}. One should also mention in this
context the generalized versions of SM, formulated on the compact
manifolds as two-sphere~\cite{jaye,bass}, or torus~\cite{aza} instead of
the flat space as well as the light-cone formulation~\cite{mcc}.
 
Although a number of papers have already been devoted to the 
investigation of propagators in SM, a relatively small interest,
up to our knowledge, has been paid to higher order Green
functions~\cite{br}. In this paper we plan to fill up the gap with the
particular interest paid to the 4-point functions. 

In the following sections we show how they can systematically be found. In
Section~\ref{sec:sect2} we consider the Ward identities in the momentum
space and show how the infinite set of Dyson-Schwinger equations can be
reduced to only one, fully solvable equation. Particularly simple solution
is given in Section~\ref{sec:sect2:subsec:2b2f} for the function
corresponding to ``Compton scattering''. For the 4-fermion Green function
we derive in Section~\ref{sec:sect2:subsec:4f} the integral equation which
has a closed form (it does not contain any higher Green functions).  In
Section~\ref{sec:sect3} we consider the same question in the coordinate
space. Following Schwinger in the quoted work~\cite{jsch}, we find
explicit solutions for both 4-point Green functions: the 4-fermion and the
2-photon-2-fermion one. Both of them are expressible through the known
scalar factors of the fermion propagator. For the most interesting case of
four fermions we use the derived formula to show that the function
contains a pole at $p^2=\mu^2$, i.e. corresponding to the Schwinger
boson. We also give a formula for the formfactor of the appropriate
residue. In the appendix we give the definitions of all the Green
functions considered in the present work.
 
\section{Momentum space 4-point Green functions} 
\label{sec:sect2} 
 
In this section we are concentrating on the momentum space equations for
the 4-point Green functions. Firstly, we deal with 2-fermion-2-boson
function. We recapitulate Ward identities which allow us to represent it
as the appropriate combination of the 3-point functions. These, however,
are already known and expressible, once again due to Ward identities,
through the full fermion propagator~\cite{thza}.

Secondly, we consider 4-fermion Green function. In this case the situation
is much more difficult since we do not have at our disposal any identity
which would permit to reduce the problem to lower functions. Therefore, we
consider the Dyson-Schwinger equation which couples the 4-point function
to 5-point one (with one boson and four fermion `legs'). Next, we apply
both Ward identities to the latter, and in consequence obtain a
selfconsistent integral equation which contains only the 4-fermion
function (and lower ones).
 
\subsection{Notation and Definitions} 
\label{sec:sect2:subsec:def} 
 
SM may be defined through the two-dimensional lagrangian density  
\begin{equation} 
{\cal L}(x)=\overline{\Psi}(x)\left(i\gamma^{\mu}\partial_{\mu} -
eA^{\mu}(x)\gamma_{\mu}\right)\Psi (x)-
\frac{1}{4}F^{\mu\nu}(x)F_{\mu\nu}(x)-
\frac{\lambda}{2}\left(\partial_{\mu}A^{\mu}(x)\right)^2\; , 
\label{lagr} 
\end{equation} 
where $\lambda$ is the gauge fixing parameter. For our calculations it
will be convenient to choose later the Landau gauge by setting 
$\lambda\rightarrow\infty$. 
For the Dirac gamma matrices the following convention will be used 
$$ 
\gamma^0=\left(\begin{array}{lr}0 & \hspace*{2ex}1 \\ 1 & 0 
\end{array}\right)\; , \;\;\;\;\; 
\gamma^1=\left(\begin{array}{lr} 0 & -1 \\ 1 & 0 
\end{array}\right)\; , \;\;\;\;\; 
\gamma^5=\gamma^0\gamma^1=\left( 
\begin{array}{lr} 1 & 0 \\ 0 & -1 \end{array}\right) \; ,
$$ 
and for the metric tensor 
$$
g^{00}=-g^{11}=1\; . 
$$ 
The totally antisymmetric symbol $\varepsilon^{\mu\nu}$ is defined by 
$$
\varepsilon^{01}=-\varepsilon^{10}=1\; ,\;\;\;\;\; 
\varepsilon^{00}=\varepsilon^{11}=0\; . 
$$

The definitions of all the Green function that appear in the formulae
below are collected together in the appendix.

\subsection{Calculation of the 2-boson, and 2-fermion functions} 
\label{sec:sect2:subsec:2b2f} 
 
We start this section with derivig Ward identities satisfied by the 
relevant Green function. The standard procedure in this derivation is to 
perform, under the functional integral~(\ref{gener}), 
the following infintesimal local gauge transformation 
\begin{eqnarray} 
A^{\mu}(x)&&\rightarrow A^{\mu}(x) + 
\partial^{\mu}\omega(x)\; ,\nonumber\\ 
\Psi(x)&&\rightarrow \Psi(x) - ie\omega(x)\Psi(x)\, 
,\label{gauge1}\\ 
\overline{\Psi}(x)&&\rightarrow \overline{\Psi}(x) + 
ie\omega(x)\overline{\Psi}(x)\; ,\nonumber 
\end{eqnarray}  
and consider the resulting variational equation.
Doing that way we get 
the relation satisfied by the generating functional 
$W(\eta,\overline{\eta},J)$ 
\begin{equation} 
-\lambda\Box_x\partial^{\mu}_x\frac{\delta W}{\delta 
J^{\mu}(x)}-\partial^{\mu}_xJ_{\mu}(x) - 
ie\overline{\eta}_a(x)\frac{\delta 
W}{\delta\overline{\eta}_a(x)} + ie\eta_a(x)\frac{\delta 
W}{\delta\eta_a(x)} = 0 
\label{infgen1} 
\end{equation} 
Now we have to functionally differentiate both sides of this 
equation over $J^{\nu}(y)$, $\overline{\eta}_b(z)$, and 
$\eta_c(u)$. After having put all the external currents at 
zero value we obtain the following equation for the 4-point Green 
function $\Gamma^{\mu\nu}$, defined in the appendix,
\begin{eqnarray} 
&&i\lambda\Box_x\partial_x^{\mu}\int 
d^2w_1d^2w_2d^2w_3d^2w_4D_{\mu\alpha}(x-w_1)S(z-
w_3)\Gamma^{\alpha\beta}(w_1,w_2;w_3,w_4)S(w_4-u)\times\nonumber\\ 
&&D_{\beta\nu}(w_2-y)= -ie^2\int 
d^2w_1d^2w_2d^2w_3S(z-
w_2)\Gamma^{\alpha}(w_1;w_2,w_3)S(w_3-
u)D_{\alpha\nu}(w_1-y)\times\nonumber\\ 
&&\left[\delta^{(2)}(x-z)-\delta^{(2)}(x-u)\right]\; , \label{a1}
\end{eqnarray} 
where we omitted the obvious spinor indices. If we now 
make use of the well known~\cite{ramond} Ward identity 
satisfied by the photon propagator 
\begin{equation} 
\lambda\Box_x\partial_x^{\mu}D_{\mu\nu}(x-y) = 
\partial^x_{\nu}\delta^{(2)}(x-y)\; ,
\label{wardphot1} 
\end{equation}
which stresses that only the transverse part of $D^{\mu\nu}$ is influenced
by the interaction,
and rewrite the expression in momentum space using the 
definitions of Figure~\ref{mom}, we obtain, after
removing the common factors on both sides, 
\begin{eqnarray} 
ik_{\mu}S(p+q-k)\Gamma^{\mu\nu}(k,q,p)S(p)&& = 
e^2\bigg[S(p+q)\Gamma^{\nu}(q,p)S(p)\nonumber\\ &&- 
S(p+q-k)\Gamma^{\nu}(q,p-k)S(p-k)\bigg]\; .\label{ward} 
\end{eqnarray} 

Obviously, this equation does not define $\Gamma^{\mu\nu}$
entirely, but only its longitudinal part (in index $\mu$). 
Fortunately, due to the vanishing electron mass the Lagrangian $\cal L$ is 
invariant also with respect to the local {\em chiral} gauge 
transformation. In the infinitesimal version they read 
\begin{eqnarray} 
A^{\mu}(x)&&\rightarrow 
A^{\mu}(x)+\varepsilon^{\mu\nu}\partial_{\nu}\omega(x)\; 
,\nonumber\\  
\Psi(x)&&\rightarrow \Psi(x)-ie\omega(x)\gamma^5\Psi(x)\; 
,\label{gauge2}\\ 
\overline{\Psi}(x)&&\rightarrow\overline{\Psi}(x)-
ie\omega(x)\overline{\Psi}(x)\gamma^5\; .\nonumber 
\end{eqnarray} 
 Similarly as it was done to obtain~(\ref{infgen1}) we can 
derive the following equation for the generating functional $W$
\begin{equation} 
\left(\Box_x+\frac{e^2}{\pi}\right)\varepsilon^{\mu\nu}
\partial^x_{\mu}\frac{\delta W}{\delta J^{\nu}(x)}-
\varepsilon^{\mu\nu}\partial^x_{\nu}J_{\mu}(x) - 
ie\overline{\eta}_a(x)\gamma_5^{ab}\frac{\delta 
W}{\delta\overline{\eta}_b(x)} - 
ie\eta_a(x)\gamma_5^{ba}\frac{\delta W}{\delta\eta_b(x)} = 
0\; . 
\label{infgen2} 
\end{equation} 
One important difference in comparison with 
eq.~(\ref{infgen1}), which should be noted here, is the presence 
of the mass equal to $\frac{e^2}{\pi}$ in the first term. This term 
results, as mentioned in the Introduction, from the 
noninvariance of  the path integral measure with respect to the 
group of transformations~(\ref{gauge1}) and constitutes the well
known chiral anomaly~\cite{kijo,rosk,abh}. Following the 
same way as above, and using the chiral version of 
equation~(\ref{wardphot1}) 
\begin{equation} 
\left( 
\Box_x+\frac{e^2}{\pi}\right)\varepsilon^{\mu\nu}\partial^x_
{\mu}D_{\nu\alpha}(x-y) = -
\varepsilon_{\alpha\nu}\partial_x^{\nu}\delta^{(2)}(x-y)\; , 
\label{wardphot2} 
\end{equation} 
we get in momentum space 
\begin{eqnarray} 
i\varepsilon_{\mu\alpha}k^{\alpha}S(p+q-
k)\Gamma^{\mu\nu}(k,q,p)S(p)&& = 
e^2\bigg[\gamma^5S(p+q)\Gamma^{\nu}(q,p)S(p)\nonumber\\ 
&&+ S(p+q-k)\Gamma^{\nu}(q,p-k)S(p-k)\gamma^5\bigg]\; 
.\label{wardchir} 
\end{eqnarray} 
This, together with~(\ref{ward}), determines uniquely $\Gamma^{\mu\nu}$
because in two dimensions there are only 
two independent space-time 2-vectors. Therefore we can write 
\begin{equation} 
g^{\mu\nu}=\frac{1}{k^2}\left(k^{\mu}k^{\nu}-
\varepsilon^{\mu\alpha}k_{\alpha}\varepsilon^{\nu\beta}k_{\beta}\right)\;
, 
\label{tens} 
\end{equation} 
and in consequence, any tensor $A^{\mu\nu}$ may be written 
as 
\begin{equation} 
A^{\mu\nu}=g^{\mu\lambda}A_{\lambda}^{\nu}=\frac{k^{\mu}}
{k^2}\left(k_{\lambda}A^{\lambda\nu}\right)-
\frac{\varepsilon^{\mu\alpha}k_{\alpha}}{k^2}\left(
\varepsilon_{\lambda\beta}k^{\beta}A^{\lambda\nu}\right)\; . 
\label{atens} 
\end{equation} 
Applying this to the 4-point function $\Gamma^{\mu\nu}$, we find 
\begin{eqnarray} 
S(p+q-k)\Gamma^{\mu\nu}(k,q,p)S(p)&&= -
\frac{ie^2}{k^2}\bigg[\not\! 
k\gamma^{\mu}S(p+q)\Gamma^{\nu}(q,p)S(p)\nonumber\\ 
&&-S(p+q-k)\Gamma^{\nu}(q,p-k) \not\! 
k\gamma^{\mu}S(p-k)\bigg]\; . 
\label{gmn} 
\end{eqnarray} 
In deriving this equation we made use of the fact, that 
$k^{\mu}-\varepsilon^{\mu\alpha}k_{\alpha}\gamma^5=\not\! 
k\gamma^{\mu}$, as well as that the propagator $S$ is linear 
in gamma matrices~\cite{jsch}, and consequently 
$\{S,\gamma^5\}=0$ (at least in the 0-instanton sector to which we
restrict ourselves in the present paper). In that way the 4-point 
function is given in terms of the vertex function and the
propagator. The vertex function, however, thanks to the 
analogous Ward identities, can be further reduced~\cite{stam,thza} to
the form
\begin{equation} 
\Gamma^{\nu}(q,p)=\frac{1}{q^2}\left[S^{-
1}(p+q) - S^{-1}(p)\right]\not\! q\gamma^{\nu}\; . 
\label{gn} 
\end{equation} 
Applying this, we obtain our final equation for the 4-point Green function
\begin{eqnarray} 
&&S(p+q-k)\Gamma^{\mu\nu}(k,q,p)S(p)=\nonumber\\ 
&&-\frac{ie^2}{k^2q^2}\not\! k\gamma^{\mu}\left[S(p)-
S(p+q)-S(p-k)+S(p+q-k)\right]\gamma^{\nu}\not\! q\; 
.\label{g2mn} 
\end{eqnarray} 
From Eq.~(\ref{g2mn}) we see that the 4-point function is entirely
expressible through 2-point functions which are already known. In the same
way one can reduce to fermion propagators, by successively applying both
Ward identities, any Green function with two fermion and $n$ boson `legs'.

A different approach one 
has to make use of while dealing with functions with more than 
two fermions, since only the photon `legs' may be removed the above  way. We 
consider this question in the following paragraph. 

The simple 
structure of $\Gamma^{\mu\nu}(k,p,q)$ reflects the fact that 
in the Schwinger Model the external photons (i.e. 
nonperturbative, massive photons) are always coupled directly 
to the electron line, without an intermediate fermion loop, 
since any loop with more than two external photons turns out 
to be zero, if we consider all possible permutations of vertices. 
 
\subsection{Selfconsistent equation for the 4-fermion function} 
\label{sec:sect2:subsec:4f} 
 
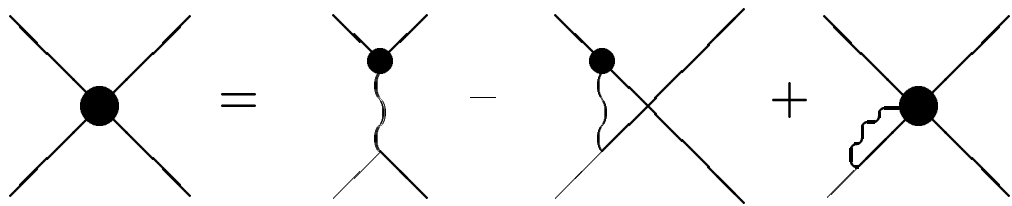
\begin{figure}[p] 
\caption{Graphic representation of the Dyson-Schwinger 
equation for the 4-point Green function: 
$\Gamma_{ab;cd}$. Thick lines correspond to full propagators and thin to
free ones.} 
\label{dys4} 
\end{figure}

In the case of the 4-fermion function the exploiting of the ordinary, 
and the chiral Ward identity does not solve the problem 
completely (although it is still very useful), since we cannot 
reduce fermion legs, in any way. However, we are able to obtain a
selfconsistent equation for this function. To do so we start from the 
Dyson-Schwinger equation for the 4-fermion function which can 
be derived in a standard way. We consider the functional 
derivative over field $\overline{\Psi}(x)$~\cite{iz}, and write  
\begin{equation} 
\int D\Psi D\overline{\Psi} 
DA\frac{\delta}{\delta\overline{\Psi}(x)}e^{i\int 
d^2x\left[{\cal 
L}(x)+\overline{\eta}(x)\Psi(x)+\overline{\Psi}(x) 
\eta(x)+J^{\mu}(x) A_{\mu}(x)\right]}=0\; . 
\label{deriv} 
\end{equation} 
This gives the following relation for the generating functional 
\begin{equation} 
i\gamma^{\mu}_{ab}\partial^{\mu}_x\frac{\delta 
W}{\delta\overline{\eta}_b(x)}-e\frac{\delta W}{\delta 
J^{\mu}(x)}\gamma^{\mu}_{ab}\frac{\delta 
W}{\delta\overline{\eta}_b(x)}+ie\gamma^{\mu}_{ab}\frac{\delta^2W}{\delta 
J^{\mu}(x)\delta\overline{\eta}_b(x)}+\eta_a(x)=0\; . 
\label{dysgen} 
\end{equation} 
Now, one has to differentiate the above equation over the fermionic currents 
$\eta_d(w)$, $\eta_c(z)$, and $\overline{\eta}_e(y)$, and at the end
set all the currents $\eta$, $\overline{\eta}$, $J$ to be equal zero.
The result is 
\begin{eqnarray} 
&&-i\gamma^{\mu}_{ab}\partial_{\mu}^x(-i)\int 
d^2w_1d^2w_2d^2w_3d^2w_4 S_{bf}(x-w_1)S_{eg}(y-
w_2)\Gamma_{fg;rs}(w_1,w_2;w_3,w_4)\times\nonumber\\ 
&&S_{rc}(w_3-z)S_{sd}(w_4-w)=e^2\int 
d^2w_1d^2w_2d^2w_3D_{\mu\nu}(x-
w_1)S_{ef}(y-w_2)\Gamma^{\nu}_{fg}(w_1;w_2,w_3)\times\nonumber\\ 
&&S_{gc}(w_3-z)\gamma^{\mu}_{ab}S_{bd}(x-w) -e^2\int 
d^2w_1d^2w_2d^2w_3D_{\mu\nu}(x-
w_1)S_{ef}(y-w_2)\Gamma^{\nu}_{fg}(w_1;w_2,w_3)\times \nonumber\\  
&&S_{gd}(w_3-w)\gamma^{\mu}_{ab}S_{bc}(x-
z)+ie\gamma^{\mu}_{ab}\int 
d^2w_1d^2w_2d^2w_3d^2w_4d^2w_5D_{\mu\nu}(x-
w_1)\times\label{dscoor}\\ 
&&S_{bf}(x-w_2)S_{eg}(y-
w_3)\Gamma^{\nu}_{fg;rs}(w_1;w_2,w_3;w_4,w_5) 
S_{rc}(w_4-z)S_{sd}(w_5-w)\; .\nonumber 
\end{eqnarray} 
A schematical representation of this equation is shown on 
Figure~\ref{dys4}. There the thick lines correspond to full propagators 
and thin lines to the free ones. This equation indicates that the 
4-point function depends on the 5-point one which is, of 
course, the typical behaviour of the set of Dyson-Schwinger 
equations, since the interacting Lagrangian contains always terms of at
least third order in fields, resulting in an infinite interlacement of
Green functions. We can, however, use here the method of previous 
paragraph and get rid of the 5-point function from the right 
hand side. Equation~(\ref{dscoor}), transformed into the 
momentum space according to the definitions of 
Figure~\ref{mom}, reads  
\begin{eqnarray} 
&&i\not\! p_{ab}S_{bf}(p) S_{eg}(q+l-
p)\Gamma_{fg;rs}(p,q,l)S_{rc}(q)S_{sd}(l) =
e^2S_{ef}(q+l-p)\Gamma^{\nu}_{fg}(l-
p,q)\times \nonumber\\ 
&&S_{gc}(q)D_{\mu\nu}(p-l)\gamma^{\mu}_{ab}S_{bd}(l) 
-e^2 S_{ef}(q+l-p)\Gamma^{\nu}_{fg}(q-
p,l)S_{gd}(l)D_{\mu\nu}(p-
q)\gamma^{\mu}_{ab}S_{bc}(q)\nonumber\\ 
&&+ie\gamma^{\mu}_{ab}\int\frac{d^2k}{(2\pi)^2}D_{\mu\nu}
(k)S_{bf}(p-k)S_{eg}(q+l-p)\Gamma^{\nu}_{fg;rs}(k,p-
k,q,l)S_{rc}(q)S_{sd}(l)\; . \label{dsmom} 
\end{eqnarray} 
Since the vertices $\Gamma^{\mu}_{ab}$ are 
perfectly known, what we need is only the relation which would 
allow us to express the 5-point function 
$\Gamma^{\mu}_{ab;cd}$ through the 4-point one. The 
method is analogous to that shown in detail in the previous 
paragraph, and we do not repeat it here. The result is 
\begin{eqnarray} 
S(p)\otimes &&S(q+l-p-k)\cdot\Gamma^{\nu}(k,p,q,l)\cdot S(p)\otimes 
S(l)=\label{five}\\  
=&&-\frac{ie}{k^2}\bigg[\left(\not\! k 
\gamma^{\nu}S(p+k)\right)\otimes S(q+l-p-
k)\cdot\Gamma(p+k,q,l)\cdot S(q)\otimes S(l)\nonumber\\ 
&&+S(p)\otimes\left(\not\! k\gamma^{\nu}S(q+l-p)\right) 
\cdot \Gamma(p,q,l)\cdot S(q)\otimes S(l)\nonumber\\ &&-S(p)\otimes 
S(q+l-p-k)\cdot \Gamma(p,q,l-k)\cdot S(q)\otimes \left(\not\! 
k\gamma^{\nu}S(l-k)\right)\nonumber\\ &&-S(p)\otimes 
S(q+l-p-k)\cdot \Gamma(p,q-k,l)\cdot \left(\not\! k\gamma^{\nu}S(q-k) 
\right)\otimes S(l)\bigg]\; ,\nonumber 
\end{eqnarray}  
where, for abbreviation, we have used the obvious notation 
$S^{(1)}\otimes S^{(2)}\cdot \Gamma\cdot S^{(3)}\otimes S^{(4)}$ 
for defining an object: $S^{(1)}_{ae}S^{(2)}_{bf} \Gamma_{ef;gh} 
S^{(3)}_{gc}S^{(4)}_{hd}$. 
If we substitute~(\ref{five}), together with~(\ref{gn}), 
into~(\ref{dsmom}) we obtain the final equation 
\begin{eqnarray} 
i\not\! p_{ab}S_{bf}(p)&&S_{eg}(q+l-
p)\Gamma_{fg;rs}(p,q,l) S_{rc}(q)S_{sd}(l) = 
\label{solfive}\\ 
&&=\frac{e^2}{(l-p)^2}\left[(\not\! l-\not\! p)\gamma^{\nu} 
\left(S(q)-S(q+l-p)\right)\right]_{ec}D_{\mu\nu}(p-
l)\gamma^{\mu}_{ab} S_{bd}(l) \nonumber\\ 
&&-\frac{e^2}{(q-p)^2}\left[(\not\! q-\not\! 
p)\gamma^{\nu}\left(S(l)-S(q+l-
p)\right)\right]_{ed}D_{\mu\nu}(p-q) 
\gamma^{\mu}_{ab}S_{bc}(q) \nonumber\\ 
&&+e^2\gamma^{\mu}_{ab}\int\frac{d^2k}{(2\pi)^2}\frac{
D_{\mu\nu}(k)}{k^2} \bigg[\left(\not\! 
k\gamma^{\nu}S(p)\right)_{bf}S(q+l-p)_{eg} 
\Gamma_{fg;rs}(p,q,l)S_{rc}(q)S_{sd}(l)\nonumber\\ 
&&+S(p-k)_{bf}\left(\not\! k\gamma^{\nu}S(q+l-p+k) 
\right)_{eg} 
\Gamma_{fg;rs}(p-k,q,l)S_{rc}(q)S_{sd}(l)\nonumber\\ 
&&-S(p-k)_{bf}S(q+l-p)_{eg}\Gamma_{fg;rs}(p-k,q,l-
k)S_{rc}(q)  \left(\not\! k\gamma^{\nu}S(l-
k)\right)_{sd}\nonumber\\ 
&&-S(p-k)_{bf}S(q+l-p)_{eg}\Gamma_{fg;rs}(p-k,q-k,l)  
\left(\not\! k\gamma^{\nu}S(q-k) \right)_{rc} S(l)_{sd}\; ,\nonumber 
\end{eqnarray} 
in which only the 4- and 2-point functions are involved.

This derivation shows how the infinite series of coupled 
Dyson-Schwinger equations may be reduced to only one 
integral equation, which in principle might be solved. 
Due to the complicated tensor structure of $\Gamma_{ab;cd}$
(it requires introducing several scalar coefficient function), 
and perplexed mathematical form of~(\ref{solfive}) we do not try 
to solve it here and will rather 
concentrate on finding the explicit form of the 4-fermion Green 
function in coordinate space. 

It is a common feature of the Schwinger
Model that coordinate space solutions are much simpler than the
momentum space ones. 
In this case it will be even possible to 
express $\Gamma_{ab;cd}$ through the electron propagator,
similarly as it was done for $\Gamma^{\mu\nu}$ in~(\ref{g2mn}). This
problem will constitute the subject of the next section. 

The method of the present paragraph allows also to find a selconsistent
equation for any higher Green function. In particular, a function with
$2n_f$ fermionic legs and $n_b$ bosonic legs should first be reduced,
thanks to the consecutive $n_b$ applications of both Ward identities, to
purely fermionic $2n_f$-point function, and than the selfconsistent
equation for the latter can be obtained.  
 
\section{4-point Green functions in coordinate space} 
\label{sec:sect3} 
 
In this section we find the explicit formulae for the 4-point
Green functions in coordinate space. In this case also 
the 4-fermion function may be given a compact form, instead of having
it as a solution of an integral equation like~(\ref{solfive}). Below we 
follow the way somehow similar to that of the original 
Schwinger's work~\cite{jsch}, but extend it also for higher 
functions. 

\subsection{2-boson and 2-fermion function}
\label{sec:sect3:subsec:2b2fc}

As it is known the generating functional 
$Z(\eta,\overline{\eta},J)$ may be given the following form  
\begin{eqnarray} 
Z(\eta,\overline{\eta},J)=&&\exp\left[-i\int 
d^2xd^2y\overline{\eta}(x){\cal S}(x,y;\delta/i\delta 
J)\eta(y)\right]\times \label{genz}\\ 
&&\exp\left[-\frac{i}{2}\int 
d^2xd^2yJ_{\mu}(x)\bigtriangleup^{\mu\nu}(x-
y;e^2/\pi)J_{\nu}(y)\right]\; ,\nonumber 
\end{eqnarray} 
where ${\cal S}(x,y, {\cal A})$ is the classical electron 
propagator in the external electromagnetic field ${\cal A}^{\mu}$, and 
is given by the formula 
\begin{equation} 
{\cal S}(x,y,{\cal A})={\cal S}_0(x-
y)\exp\left[-i\left(\tilde{\phi} (x,{\cal A})-(\tilde{\phi} (y,{\cal 
A})\right)\right]\; ,
\label{grfree} 
\end{equation}  
with 
\begin{equation} 
\tilde{\phi}(x,{\cal A})=e\int d^2y\bigtriangleup(x-
y)\gamma^{\nu}\gamma^{\mu}\partial_{\mu}{\cal 
A}_{\nu}(y)\; ,
\label{phi} 
\end{equation} 
${\cal S}_0$ being the free propagator. In the Landau gauge 
which we now use, $\bigtriangleup^{\mu\nu}$ takes the form 
\begin{equation} 
\bigtriangleup^{\mu\nu}(x-y,m^2)=\int 
d^2z\left[g^{\mu\nu}\delta^{(2)}(x-z)-
\partial^{\mu}_x\partial^{\nu}_x\bigtriangleup(x-
z)\right]\bigtriangleup(z-y;m^2)\; , 
\label{deftri} 
\end{equation} 
where $\bigtriangleup(x,m^2)$ and $\bigtriangleup(x)$ are 
respectively Klein-Gordon and d'Alambert propagators. For 
the exponential factor in~(\ref{grfree}) one often uses the 
abbreviated and useful form 
\begin{equation} 
\tilde{\phi} (x,{\cal A})-\tilde{\phi} (y,{\cal A})=-\int 
d^2z{\cal A}^{\mu}(z){\cal J}_{\mu}(z;x,y)\; , 
\label{expon} 
\end{equation} 
with the (nonconserved) current ${\cal J}^{\mu}$ stisfying 
\begin{equation} 
\partial^{\mu}_z{\cal J}_{\mu}(z;x,y)=e\left[\delta^{(2)}(x-
z)-\delta^{(2)}(y-z)\right]\; . 
\label{noncur} 
\end{equation}
This current has sources at every point, where charged particles are
created, or anihilated, and is closely related with the notion of th so called
``compensating current''~\cite{ibzb}.
Now the 4-point Green function, considered in 
Section~\ref{sec:sect2:subsec:2b2f} (for  functions with no external `legs' 
amputated we reserve symbol $G$ with appropriate indices), is given by 
\begin{equation} 
G^{\mu\nu}_{ab}(x_1,x_2;x_3,x_4)=\left.\frac{\delta^4}{\delta 
J_{\mu}(x_1)\delta 
J_{\nu}(x_2)\delta\overline{\eta}_a(x_3)\delta\eta_b(x_4)}Z
(\overline{\eta},\eta, J)\right|^{\rm connected}_{{\rm curr.}=0}\; , 
\label{g22} 
\end{equation} 
where we have explicitely written that only connected graphs are
considered. This gives 
\begin{eqnarray} 
&&G^{\mu\nu}_{ab}(x_1,x_2;x_3,x_4)= 
i\frac{\delta^2}{\delta J_{\mu}(x_1)\delta J_{\nu}(x_2)}{\cal 
S}_{ab}\left(x_3,x_4;\delta/i\delta J\right)\label{gr22}\\
&&\exp\left[-\frac{i}{2}\int d^2xd^2y 
J^{\alpha}(x)\bigtriangleup_{\alpha\beta}(x-
y;e^2/\pi)J^{\beta}(y)\right]\Bigg|^{\rm connected}_{J=0}\; .\nonumber
\end{eqnarray} 
If we make use of the explicit form of $\cal S$, given in 
equations~(\ref{grfree},\ref{phi}), leading to the representation in the
form of a series of derivatives, and note that $\exp\left(z 
\frac{d}{dx}\right) f(x) = f(x+z)$, we can write 
\begin{eqnarray} 
&&G^{\mu\nu}_{ab}(x_1,x_2;x_3,x_4)= 
i\frac{\delta^2}{\delta J_{\mu}(x_1)\delta J_{\nu}(x_2)}{\cal 
S}_0^{ac}(x_3-x_4)
\exp\bigg[-\frac{i}{2}\int d^2xd^2y\bigg(J^{\alpha}(x)+ \label{gr3}\\ 
&&{\cal J}^{\alpha}(x;x_3,x_4)\bigg)\bigtriangleup_{\alpha\beta}(x-
y;e^2/\pi) \bigg(J^{\beta}(y)+{\cal 
J}^{\beta}(y;x_3,x_4)\bigg)\bigg]_{cb}\Bigg|^{\rm 
connected}_{J=0}\; .\nonumber 
\end{eqnarray} 
Since the currents $\cal J$ defined in~(\ref{expon}) have the 
matrix structure of the form $A+B\gamma^5$, which means that they 
commute with each other, the differentiation may be easily 
performed, and we obtain 
\begin{eqnarray}
&&G^{\mu\nu}_{ab}(x_1,x_2;x_3,x_4)={\cal
S}_0^{ac}(x_3-x_4)\bigtriangleup^{\mu\nu}(x_1-
x_2; e^2/\pi)\times \label{gr4}\\
&&\exp\left[-\frac{i}{2}\int d^2xd^2y{\cal 
J}^{\alpha}(x;x_3,x_4) \bigtriangleup_{\alpha\beta}(x-
y;e^2/\pi){\cal J}^{\beta}(y;x_3,x_4)\right]_{cb}\nonumber\\
&&-i{\cal S}_0^{ac}(x_3-x_4)\int 
d^2zd^2w\bigtriangleup_{\mu\lambda}(x_1-w;e^2/\pi){\cal 
J}^{\lambda}_{cd}(w;x_3,x_4) \bigtriangleup_{\nu\rho}(x_2-
z;e^2/\pi)\times\nonumber\\
&&{\cal J}^{\rho}_{de}(z;x_3,x_4)\exp\left[-\frac{i}{2}\int d^2xd^2y{\cal 
J}^{\alpha}(x;x_3,x_4) ) \bigtriangleup_{\alpha\beta}(x-
y;e^2/\pi){\cal J}^{\beta}_{de}(y;x_3,x_4)\right]_{eb}\Bigg|^{\rm 
connected}\; .\nonumber
\end{eqnarray}
Now we recall that the full propagator $S$ has the form~\cite{jsch} 
\begin{equation}
S_{ab}(u-w)=
{\cal S}_0^{ac}(u-w)\exp\left[-\frac{i}{2}\int d^2xd^2y{\cal 
J}^{\alpha}(x;u,v) \bigtriangleup_{\alpha\beta}(x-
y;e^2/\pi){\cal J}^{\beta}(y;u,v)\right]_{cb}\; .
\label{fls}
\end{equation}
This allows us to write equation~(\ref{gr4}) in the form
\begin{eqnarray}
&&G^{\mu\nu}_{ab}(x_1,x_2;x_3,x_4)=S_0^{ab}(x_3,x_4)
\bigtriangleup^{\mu\nu}(x_1-x_2; e^2/\pi)-iS_{ac}(x_3,x_4)\label{gr5}\\
&&\int d^2zd^2w\bigtriangleup_{\mu\lambda}(x_1-w;e^2/\pi){\cal 
J}^{\lambda}_{cd}(w;x_3,x_4) \bigtriangleup_{\nu\rho}(x_2-
z;e^2/\pi){\cal J}^{\rho}_{de}(z;x_3,x_4)\Bigg|^{\rm 
connected}\; .\nonumber
\end{eqnarray}
The first term, constituting the 
nonconnected contribution, should now be rejected. For the amputated Green 
function $\Gamma$ considered in Section~\ref{sec:sect2:subsec:2b2f},
with the use of the definition of $\cal J$ and $\tilde{\phi}$, as well 
as the fact that $S\gamma^{\mu}\gamma^{\nu} = 
\gamma^{\nu}\gamma^{\mu}S$, we obtain
\begin{eqnarray}
&&\int d^2ud^2w S(x_3-
u)\Gamma^{\mu\nu}(x_1,x_2;u,w)S(w-x_4)=\label{grkon}\\
&&=-ie^2\not\!\partial_{x_1}\left(\bigtriangleup(x_3-x_1)- 
\bigtriangleup(x_4-x_1)\right)\gamma^{\mu}S(x_3-
x_4)\gamma^{\nu}\not\!\partial_{x_2} 
\left(\bigtriangleup(x_3-x_2)- \bigtriangleup(x_4-x_2)\right)\; 
,\nonumber
\end{eqnarray}
where we have omitted the spinor indices. One can easily 
verify that this is the coordinate space representation 
of~(\ref{g2mn}). It is a common feature of all two-fermion 
Green functions that they can be represented through 
fermion propagator in both coordinate and momentum spaces. 
As it was mentioned this is possible since the external 
photons are coupled directly to the incoming and 
outgoing electron lines, and no intermediate fermions loops are possible,
apart from those producing the photon mass $e^2/\pi$. On the 
other hand, for the 4-fermion function, we deal with in the 
following section, a much more complicated structure appears and 
an explicit and compact expression is possible to be given only 
in coordinate space. 

\subsection{4-fermion function}
\label{sec:sect3:subsec:4fc}

Now, instead of~(\ref{g22}), we have
\begin{eqnarray}
&&G_{ab;cd}(x_1,x_2;x_3,x_4)=
\left.\frac{\delta^4}{\delta\overline{\eta}_a(x_1)\delta
\overline{\eta}_b(x_2)\delta\eta_c(x_3)\delta\eta_d(x_4)} Z
(\overline{\eta},\eta, J)\right|^{\rm connected}_{{\rm
curr.}=0}=\label{4fgr}\\ 
&&=\left[{\cal S}_{ac}(x_1,x_3;\delta/i\delta J) {\cal
S}_{bd}(x_2,x_4;\delta/i\delta J)
-{\cal S}_{ad}(x_1,x_4;\delta/i\delta J) {\cal
S}_{bc}(x_2,x_3;\delta/i\delta J)\right]Z(J)\Bigg|^{\rm connected}_{J=0}\;
. \nonumber
\end{eqnarray} 
The differentiations over external current $J$, hidden in propagators $\cal
S$, can be performed similarly as it was done to obtain~(\ref{gr3}),
although it must be done with greater care than before due to the tensor
structure of $\cal J$. In particular, using the notation of
paragraph~\ref{sec:sect2:subsec:4f} we find 
\begin{eqnarray}
&&{\cal S}(x_1,x_3;\delta/i\delta J)\otimes {\cal S}(x_2,x_4;\delta/i\delta
J)\cdot Z(J)=\label{part}\\ 
&&={\cal S}_0(x_1-x_3)\otimes {\cal S}_0(x_2-x_4)\cdot Z\left({\bf
1}\otimes {\bf 1}\cdot J+{\cal J}(x_1,x_3)\otimes {\bf 1}+{\bf 1}\otimes{\cal
J}(x_2,x_4)\right)\; ,\nonumber 
\end{eqnarray}
where for abbreviating we have not explicitely written the first argument
of the currents $\cal J$ over which the integration in the generating
functional $Z$ is taken (see
equations~(\ref{genz},\ref{phi},\ref{expon})). Now, we 
concentrate only on the last factor of the above expression ($Z$), which,
after setting $J=0$, takes the form 
\begin{eqnarray}
&&Z\left({\cal J}(x_1,x_3)\otimes {\bf 1}+{\bf 1}\otimes{\cal
J}(x_2,x_4)\right)=\exp\bigg[-\frac{i}{2}\int d^2xd^2y\bigg({\cal
J}^{\mu}(x;x_1,x_3)\otimes{\bf 1}\label{zet1}\\ 
&&+{\bf 1} \otimes {\cal
J}^{\mu}(x;x_2,x_4)\bigg)\bigtriangleup_{\mu\nu}(x-y;e^2/\pi) \bigg({\cal
J}^{\nu}(y;x_1,x_3)\otimes{\bf 1}+{\bf 1} \otimes {\cal
J}^{\nu}(y;x_2,x_4)\bigg)\bigg]\; .\nonumber 
\end{eqnarray}
The expressions for both ${\cal J}_{\mu}$ and $\bigtriangleup_{\mu\nu}$
are known so it is only a matter of patience to get the formula for the above
exponential. Thanks to the fact that in two dimensions
$\gamma^{\mu}\gamma^{\nu}\gamma_{\mu}=0$ the ``diagonal'' terms of the
kind ${\cal J}^{\mu}\otimes{\bf 1}\cdot\bigtriangleup_{\mu\nu}\cdot{\cal
J}^{\nu}\otimes{\bf 1}$ produce only expressions of the tensor structure
${\bf 1}\otimes {\bf 1}$, whereas mixed terms as ${\cal
J}^{\mu}\otimes{\bf 1}\cdot\bigtriangleup_{\mu\nu}\cdot{\bf 1}\otimes {\cal
J}^{\nu}$ give both ${\bf 1}\otimes {\bf 1}$ and
$\gamma^5\otimes\gamma^5$. We skip this calculation here to save the
reader's time, and give below only the final result 
\begin{eqnarray}
&&Z\left({\cal J}(x_1,x_3)\otimes {\bf 1}+{\bf 1}\otimes{\cal
J}(x_2,x_4)\right)=\frac{1}{2}({\bf 1}\otimes {\bf
1}+\gamma^5\otimes\gamma^5)\times  \label{zfin}\\ 
&&\exp\left[ie^2\left(\beta(x_1-x_2)-\beta(x_1-x_3)
-\beta(x_1-x_4)-\beta(x_2-x_3)-\beta(x_2-x_4)+\beta(x_3-x_4)\right)\right]
\nonumber\\
&&+\frac{1}{2}({\bf 1}\otimes {\bf
1}-\gamma^5\otimes\gamma^5)\exp\bigg[ie^2(-\beta(x_1-x_2)-
\beta(x_1-x_3)+\beta(x_1-x_4)+\beta(x_2-x_3)\nonumber\\
&&-\beta(x_2-x_4)-\beta(x_3-x_4))\bigg]\; ,\nonumber 
\end{eqnarray} 
where the function $\beta$ is defined by
\begin{eqnarray}
\beta(x)&&=\int\frac{d^2p}{(2\pi)^2}\left(1-e^{ipx}\right)
\frac{1}{(p^2-e^2/\pi
+i\epsilon)(p^2+i\epsilon)}=\label{beta}\\
&&\nonumber\\ 
&&=\left\{\begin{array}{ll}\frac{i}{2e^2}\left[-\frac{i\pi}{2}+\gamma_E
+\ln\sqrt{ e^2x^2/4\pi}+ 
\frac{i\pi}{2}H_0^{(1)}(\sqrt{e^2x^2/\pi})\right] & \hspace*{3ex}
x\;\;\;\; {\rm timelike}\\ 
\frac{i}{2e^2}\left[\gamma_E+\ln\sqrt{-e^2x^2/4\pi}+K_0(\sqrt{-e^2x^2/\pi})
\right] &  \hspace*{3ex}x\;\;\;\; {\rm spacelike}\; ,\end{array}\right.
\nonumber
\end{eqnarray}
and is in fact a function of $x^2$ only. $\gamma_E$ is here the Euler constant
and functions $H_0^{(1)}$ and $K_0$ are Hankel function of the
first kind, and Basset function respectively\cite{old}. Since we
have~\cite{jsch} 
\begin{equation}
S(x)={\cal S}_0(x)\exp\left[-ie^2\beta(x)\right]\; ,
\label{propbet}
\end{equation}
we can write down the final formula for the 4-fermion Green function
\begin{eqnarray}
&&G_{ab;cd}(x_1,x_2;x_3,x_4)=\frac{1}{2}\bigg[S_{ac}(x_1-x_3)S_{bd}(x_2-x_4)
+\left(S(x_1-x_3)\gamma^5\right)_{ac}\left(S(x_2-x_4)\gamma^5\right)_{bd}
\bigg]\times\nonumber\\
&&\exp\bigg[ie^2(\beta(x_1-x_2)
-\beta(x_1-x_4)-\beta(x_2-x_3)+\beta(x_3-x_4))\bigg]\nonumber\\
&&+\frac{1}{2}\bigg[S_{ac}(x_1-x_3)S_{bd}(x_2-x_4)- \left(S(x_1-x_3)\gamma^5
\right)_{ac}\left(S(x_2-x_4)\gamma^5\right)_{bd}\bigg]\nonumber\\
&&\exp\bigg[-ie^2(\beta(x_1-x_2)-\beta(x_1-x_4)-\beta(x_2-x_3)
+\beta(x_3-x_4))\bigg]\nonumber\\
&&-\left\{\begin{array}{ccc} x_3 & \longleftrightarrow & x_4\\ c &
\longleftrightarrow & d 
\end{array}\right\}\; .\label{fin4f}
\end{eqnarray}
We see that in the coordinate space both 4-point functions (4-fermion
and 2-boson-2-fermion) may perfectly be
found and are given by compact formulae. Since $\beta$'s are related with
the full fermion propagator $S$, one can say that knowing $S$ one
knows ``everything''. The calculation of higher functions may be lead
very much similarly to what was given in this paragraph, and one will
always obtain a product of electron propagators and exponentials of
$\beta$ function.

The exact expression for the 4-fermion function, we have obtained, allows
the analysis of its analytical properties. We concentrate below on the
presence of the fermion-antifermion pole ($t$-channel) corresponding to
the Schwinger boson. Let us denote the first two terms on the right hand side
of~(\ref{fin4f}) by $G^1_{ab;cd}$ and $G^2_{ab;cd}$ respectively. The
remaining terms represented by the curly brackets can contribute to the
eventual pole in the $u$-channel only, and therefore we omit them in the
present discussion.

While looking for a pole we first identify the ``in'' and ``out''
coordinates (in the $t$-channel) of fermion and antifermion:
$u\equiv x_1=x_3$, $v\equiv x_2=x_4$ and next consider the expression
Fourier transformed in the variable $z\equiv v-u$. The identification has
to be performed with care, for instance in the following way:
\begin{enumerate}
\item for the time coordinates we put 
$$
x_1^0=x_3^0\rightarrow u^0\;\;\;\; {\rm and}\;\;\;\;
x_2^0=x_4^0\rightarrow v^0\;,
$$
\item for the spacial coordinates we assume 
$$
x_1^1\rightarrow u^1\; ,\;\; x_3^1\rightarrow u^1+\varepsilon\; ,\;\;
x_2^1\rightarrow v^1\; ,\;\; x_4^1\rightarrow v^1+\eta\; ,
$$
\item for the function depending on $\varepsilon$ and $\eta$ we take the
fully symmetric limit
$$
\lim_{\stackrel{\scriptstyle\varepsilon\rightarrow 0}
{\eta\rightarrow 0}}^{sym.}f(\varepsilon,\eta) \equiv
\frac{1}{4}\lim_{\stackrel{\scriptstyle\varepsilon\rightarrow 0}
{\eta\rightarrow 0}}\left[f(\varepsilon,\eta)+f(-\varepsilon,\eta)+
f(\varepsilon,-\eta)+f(-\varepsilon,-\eta)\right]
$$
\end{enumerate}
In that limit $G^1$ and $G^2$ become only $z$ dependent. For instance for
$G^1$ we have
\begin{eqnarray}
&&G^1(z)=\frac{1}{8\pi^2}\left(\gamma^0\otimes\gamma^0 +
\gamma^1\otimes\gamma^1\right)
\lim_{\stackrel{\scriptstyle\varepsilon\rightarrow 0} 
{\eta\rightarrow 0}}^{sym.}\frac{1}{\varepsilon\eta}\times
\label{eq:g1sym}\\ 
&&\exp\left[ie^2\left(\beta(z)-\beta(0,\varepsilon)-\beta(z^0,z^1+\eta)-
\beta(z^0,z^1-\varepsilon)
-\beta(0,\eta)+\beta(z^0,z^1-\varepsilon+\eta)\right)\right]\; ,\nonumber
\end{eqnarray}
where, when it was necessary, we wrote explicitely both coefficients of the
two-vector argument of the $\beta$ function
$$
\beta(x)=\beta(-x)\equiv\beta(x^0,x^1)\; .
$$

The symmetric limit above may be performed in a strightforward way, since
the $\beta$ function is perfectly known, and we obtain
\begin{equation}
G^1(z)=-\frac{ie^2}{8\pi^2}\left(\gamma^0\otimes\gamma^0 +
\gamma^1\otimes\gamma^1\right)\frac{d^2}{dz^2}\beta(z)\; .
\label{eq:g1fin}
\end{equation}

The same limit for $G^2$ gives
\begin{equation}
G^2(z)=-\frac{ie^2}{8\pi^2}\left(\gamma^0\otimes\gamma^0 -
\gamma^1\otimes\gamma^1\right)\frac{d^2}{dz^2}\beta(z)\; .
\label{eq:g2fin}
\end{equation}

If we now apply explicitely the definition of $\beta$ given
by~(\ref{beta}), and perform the Fourier transform over $z$, we find the
following expression for the ``polar'' part of $G$
\begin{equation}
G_{polar}(k)=-\frac{ie^2}{4\pi^2}\gamma^0\otimes\gamma^0\frac{(k^1)^2}
{(k^2-e^2/\pi+i\epsilon)(k^2+i\epsilon)}\longrightarrow
-\frac{i}{4\pi}\gamma^0\otimes\gamma^0\frac{(k^1)^2}{(k^2-e^2/\pi+i\epsilon)}
\; ,
\label{eq:polar}
\end{equation}
from which a pole corresponding to the Schwinger boson may clearly be seen.

It should be noted that the similar analysis, although much more
complicated, may be done without identyfing the ``in'' and ``out''
coordinates. One can for example introduce the new c.m. variables
$$
u=\frac{1}{2}\left(x_1+x_3\right)\;,
\;\;v=\frac{1}{2}\left(x_2+x_4\right)\; ,
$$
and the relative ones
$$
x=x_1-x_3\; ,\;\; y=x_2-x_4\; .
$$
The Fourier transform of $G$ performed over $z=v-u$ displays now much
richer analytical structure (branch poins at $k^2=n^2e^2/\pi$, $n=2, 3,
...$) and the residue in the Schwinger pole depends on the relative
coordinates $x$ and $y$
\begin{equation}
G_{polar}(x,y;k)=-4i\pi\left(S(x)\gamma^5\right)\otimes
\left(S(y)\gamma^5\right)\frac{\sin \left[kx/2\right] \sin 
\left[ky/2\right]}{(k^2-e^2/\pi+i\epsilon)}\; ,
\label{eq:fullpol}
\end{equation} 
where $S$ is given by~(\ref{propbet}). For $x,y\rightarrow 0$ (in a
symmetrical way) we reproduce the result given by~(\ref{eq:polar}). It may
be noted that the formfactor $F(x)\sim S(x)\gamma^5\sin kx/2$ is square
normalizable in the sense
$\displaystyle\int_{-\infty}^{\infty}dx^1\left|F(0,x^1)\right|^2$. 

\section{Summary} 
\label{sec:sum} 
 
Below we would like to recapitulate the results we obtained in the present
work. At first, in Section~\ref{sec:sect2}, we considered Ward identities
in momentun space satisfied by the 4- and 5-point Green functions.  Thanks
to the local chiral symmetry of the Lagrangian, apart from the ordinary
gauge invariance, we derived two identities. In the 2-dimensional world
these two identities suffice to entirely describe the considered Green
function, and express it through lower ones. Each application of these
identities allows us to reduce the number of external photons by one.
Following that way we were able to reduce the 2-boson-2-fermion function
to well known electron propagator. In the case of 4-fermion function the
situation turned out to be much more severe since we have no photon legs
to reduce. The alternative approach was, therefore, introduced in
Section~\ref{sec:sect2:subsec:4f}. The starting point was here the
Dyson-Schwinger equation which, on one hand, introduces 5-point function,
but on the other permits to reduce it to the function we are looking for.
This leads to a selfconsistent integral equation which contains apart from
the unknown function only propagators which are perfectly known. We were,
unfortunately, unable to solve this integral equation because of its
complicated mathematical character, which is not unexpected since in the
Schwinger Model even the fermion propagator cannot be given an explicit
form in momentum space. The selfconsistent equation obtained in this
section may, however, be a starting point for an analysis in momentum
space constituting an alternative for taking the six-variable (two
integrations may be separated out to give the Dirac delta function)
Fourier transform.

In Section~\ref{sec:sect3} we considered the same functions in coordinate
space. We used the generating functional which had already been found in the
original Schwinger's work~\cite{jsch}. The Green functions are, of
course, given as the appropriate derivatives of this functional over the
external currents. The problem which one only has to take care of is the
tensor structure of the functions.  The final compact expressions for the
all 4-point functions were found and are shown to be expressible
through the fermion propagator. All the methods of this, as well as of
preceding section, may easily be generalized to any higher Green
functions.

For the most interesting case --- the 4-fermion function --- we were able
to show that~(\ref{fin4f}) contains a pole, in the fermion-antifermion
channel, corresponding to the Schwinger boson. It is interesting to note
that the formfactor in the residue of the pole turns out to be
normalisable in the 1-space direction if we set the relative time to
zero. However, we do
not treat this observation as any ``proof'' that the Schwinger boson is a
``bound electron-positron state'', as is here and there suggested~\cite{cks}.

\acknowledgements
The authors would like to thank very much to Dr. K. Meissner for
interesting and valuable discussions. 

\appendix
\section{Definitions of the Green functions}
\label{sec:app}

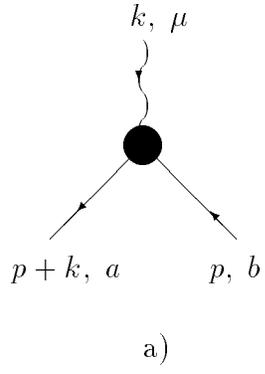
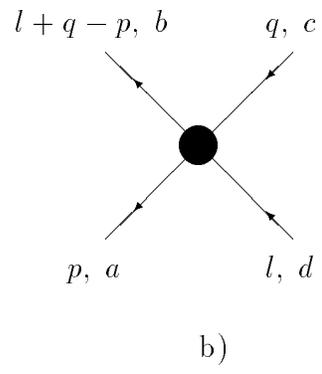
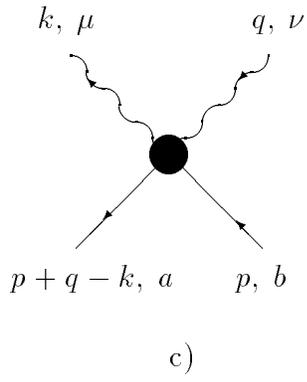
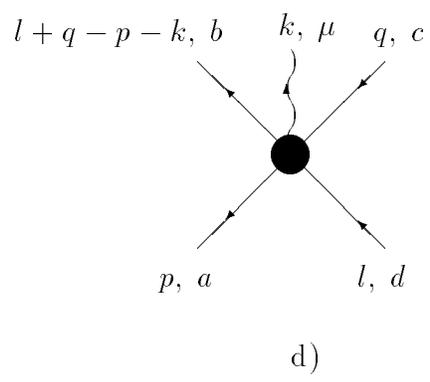
\begin{figure}[p] 
\caption{Definitions of arguments in the vertex function and 
4- and 5-point Green functions: a) 
$\Gamma^{\mu}_{ab}(k,p)$, b) $\Gamma_{ab,cd}(p,q,l)$, c) 
$\Gamma^{\mu\nu}_{ab}(k,q,p)$, d) 
$\Gamma^{\mu}_{ab;cd}(k,p,q,l)$.} 
\label{mom} 
\end{figure}

In this Appendix we give the definitions of various Green functions used
in the formulae of sections~\ref{sec:sect2} and~\ref{sec:sect3}. If we
introduce the generating functionals $Z$ and $W$ by the formula 
\begin{eqnarray} 
&&Z(\eta,\overline{\eta},J)=\exp 
iW(\eta,\overline{\eta},J)=\nonumber\\ 
&&\int D\Psi D\overline{\Psi} DA \exp\left[i\int d^2x 
\left({\cal L}(x) + \overline{\eta}(x)\Psi (x) + 
\overline{\Psi}(x)\eta (x) + 
J^{\mu}(x)A_{\mu}(x)\right)\right]\; , 
\label{gener} 
\end{eqnarray} 
we can define the connected Green functions through derivatives of the
functional over the external currents as follows 
\begin{eqnarray} 
\left.\frac{\delta^2W}{\delta\overline{\eta}_a(x)\delta\eta_b(y
)}\right|_{{\rm curr.}=0}&=&S_{ab}(x-y)\; ,\label{defgr1}\\ 
\nonumber\\ 
\left.\frac{\delta^2W}{\delta J^{\mu}(x)\delta 
J^{\nu}(y)}\right|_{{\rm curr.}=0}&=&-D_{\mu\nu}(x-y)\; 
,\label{defgr2}\\ 
\nonumber\\
\left.\frac{\delta^3W}{\delta 
J^{\mu}(x)\delta\overline{\eta}_a(y)\delta\eta_b(z)}\right|_{{\rm 
curr.}=0}&=&-e\int d^2w_1d^2w_2d^2w_3 D_{\mu\nu}(x-w_1)S_{ac}(y-
w_2)\Gamma^{\nu}_{cd}(w_1;w_2,w_3)\times\nonumber\\ 
&&S_{db}(w_3-z) \label{defgr3} \; . 
\end{eqnarray} 
We also need the 4- and 5-point functions 
\begin{eqnarray} 
\left.\frac{\delta^4W}{\delta\overline{\eta}_a(x)\delta 
\overline{\eta}_b(y)\delta\eta_c(z)\delta\eta_d(u)}\right|_{{\rm 
curr.}=0}=-i\int d^2w_1d^2w_2d^2w_3d^2w_4S_{ae}(x-
w_1)S_{bf}(y-w_2)&&\times \nonumber\\ 
\Gamma_{ef;gh}(w_1,w_2;w_3,w_4)S_{gc}(w_3-
z)S_{hd}(w_4-u)\; ,&&\label{defgr4}\\  
\nonumber\\ 
\left.\frac{\delta^4W}{\delta J^{\mu}(x)\delta 
J^{\nu}(y)\delta\overline{\eta}_a(z)\delta\eta_b(u)}\right|_{{\rm 
curr.}=0}=-i\int 
d^2w_1d^2w_2d^2w_3d^2w_4D_{\mu\alpha}(x-
w_1)S_{ac}(z-w_3)&&\times \nonumber\\ 
\Gamma^{\alpha\beta}_{cd}(w_1,w_2;w_3,w_4)S_{db}(
w_4-u)D_{\beta\nu}(w_2-y)\; ,&&\label{defgr5}\\
\nonumber\\
\left.\frac{\delta^5W}{\delta J^{\mu}(x)\delta 
\overline{\eta}_a(y)\delta\overline{\eta}_b(z)\delta\eta_c(u)\delta
\eta_d(w)}\right|_{{\rm curr.}=0}=\int 
d^2w_1d^2w_2d^2w_3d^2w_4d^2w_5D_{\mu\alpha}(x-
w_1)&&\times \nonumber\\  
S_{ae}(y-w_2)S_{bf}(z-
w_3)\Gamma^{\alpha}_{ef;gh}(w_1;w_2,w_3;w_4,w_5)S_{g
c}(w_4-u)S_{hd}(w_5-w)\; .&&\label{defgr6} 
\end{eqnarray} 
Thanks to the translational invariance of the theory these functions depend 
in fact only on the differences of arguments. The corresponding 
definitions in momentum space, after having pulled apart the 
Dirac delta function of the whole two-momentum, are given 
on Fig.~\ref{mom}.

\begin{references} 
\bibitem{jsch} J. Schwinger, in {\it Theoretical Physics},  
Trieste Lectures 1962 (I.A.E.A., Vienna 2963), p. 89; Phys. Rev. {\bf 128}, 
2425(1962). 
\bibitem{cks} A. Casher, J. Kogut and L. Susskind, Phys. Rev. 
{\bf D10}, 732(1974). 
\bibitem{iz} For instance C. Itzykson and J.-B. Zuber, {\it Quantum Field 
Theory}, (McGraw-Hill, New York 1980)  
\bibitem{fuji} K. Fujikawa, Phys. Rev. Lett. {\bf 42}, 
1195(1979); Phys. Rev. {\bf D21}, 2848(1980). 
\bibitem{rosk} R. Roskies and F. Schaposnik, Phys. Rev. {\bf 
D23}, 558(1981). 
\bibitem{sara} R. E. Gamboa Sarav\'{\i} {et al.}. Ann. Phys. 
(N.Y.) {\bf 157}, 360(1984). 
\bibitem{abh} C. Adam, R. A. Bertelmann and P. Hofer, Riv. 
N. Cim. {\bf 16}, 1(1993). 
\bibitem{stin} M. Stingl, Phys. Rev. {\bf D34}, 3863(1986). 
\bibitem{lowe} J. H. Lowenstein and J. A. Swieca, Ann. Phys. 
(N.Y.) {\bf 68}, 172(1971). 
\bibitem{jaye} C. Jayewardena, Helv. Phys. Acta {\bf 61}, 
636(1988). 
\bibitem{cad} C. Adam, Z. Phys. {\bf C63}, 169(1994).
\bibitem{cjs} S. Coleman, R. Jackiw and L. Susskind, Ann. 
Phys. (N.Y.) {\bf 93}, 267(1975). 
\bibitem{ad1} C. Adam, preprint hep-th/9704064. 
\bibitem{thir} W. E. Thirring, Ann. Phys. (N.Y.) {\bf 3}, 
91(1958). 
\bibitem{sal} A. Salam, Phys. Rev. {\bf 130}, 1287 (1963). 
\bibitem{gt} R. Delbourgo and P. West, Phys. Lett. {\bf 72B}, 
96(1977); J. Phys. {\bf A10}, 1049(1977); R. Delbourgo, N. 
Cim. {\bf 49A}, 484 (1979); R. Delbourgo, R. Zhang, J. Phys. 
{\bf A17}, 3593(1984);  C. N. Parker, J. Phys. {\bf A17}, 
2873(1984); G. Thompson, R. Zhang, Phys. Rev. {\bf D35}, 
631(1987).  
\bibitem{stam} K. Stam, J. Phys. G: Nucl. Phys. {\bf 9}, L229(1983). 
\bibitem{yild} A. Yildiz, Physica {\bf 96A}, 341(1979). 
\bibitem{wotz} C. Wotzasek, Acta Phys. Polon. {\bf B21}, 
457(1990). 
\bibitem{schm} A. U. Schmidt, Univ. Iagiell. Acta Math., 
fasc. XXXIV (1996). 
\bibitem{bass} A. Bassetto and L. Griguolo, Nucl. Phys. {\bf B439},
327(1995).
\bibitem{aza} S. Azakov, preprint hep-th/9608103.
\bibitem{mcc} G. McCartor, Z. Phys. {\bf C52}, 611(1991).
\bibitem{br} L. S. Brown, N. Cim. {\bf 29}, 617(1963). 
\bibitem{thza} G. Thompson and R. Zhang, J. Phys. G: Nucl. 
Phys. {\bf 13}, L93(1987). 
\bibitem{ramond} For instance: P. Ramond, {\it Field Theory. 
A Modern Primer}, (Benjamin/Cummings, London 1981). 
\bibitem{kijo} J. Kijowski, G. Rudolph and M. Rudolph, hep-th/9710003. 
\bibitem{ibzb} I. Bia{\l}ynicki-Birula and Z. Bia{\l}ynicka-Birula, {\it
Quantum Electrodynamics}, (Pergamon, Oxford 1975).
\bibitem{old} J. Spanier and K. B. Oldham, {\it An atlas of functions}
(HPC-Springer, Washington, Berlin 1987).

\end{references}
\end{document}